\documentclass[12pt]{article}
\pdfoutput=1                     
\usepackage{graphicx,amssymb,bbold}
\usepackage{amsmath,amsfonts,epsfig}
\usepackage{braket}
\usepackage{cite}
\usepackage[usenames]{color}

\newcommand{\pslash}{p\kern-1ex /}
\newcommand{\qslash}{q\kern-1ex /}
\newcommand{\lslash}{l\kern-1ex /}
\newcommand{\sslash}{s\kern-1ex /}
\newcommand{\kaslash}{k_a\kern-2ex /}
\newcommand{\kbslash}{k_b\kern-2ex /}
\newcommand{\Dslash}{{\cal D}\kern-1.5ex /}

\newcommand{\bc}{\overline{c}}

\newcommand{\beqa}{\begin{eqnarray}}
\newcommand{\eeqa}{\end{eqnarray}}

\newcommand{\bpm}{\begin{pmatrix}}
\newcommand{\epm}{\end{pmatrix}}
\newcommand{\bbm}{\begin{bmatrix}}
\newcommand{\ebm}{\end{bmatrix}}

\makeatletter

\@addtoreset{equation}{section}
\makeatother
\usepackage{subfig}
\begin{document}


\voffset -0.7 true cm
\hoffset 1.5 true cm
\topmargin 0.0in
\evensidemargin 0.0in
\oddsidemargin 0.0in
\textheight 8.6in
\textwidth 5.4in
\parskip 9 pt
 
\def\Tr{\hbox{Tr}}
\newcommand{\be}{\begin{equation}}
\newcommand{\ee}{\end{equation}}
\newcommand{\bea}{\begin{eqnarray}}
\newcommand{\eea}{\end{eqnarray}}
\newcommand{\beas}{\begin{eqnarray*}}
\newcommand{\eeas}{\end{eqnarray*}}
\newcommand{\nn}{\nonumber}
\font\cmsss=cmss8
\def\C{{\hbox{\cmsss C}}}
\font\cmss=cmss10
\def\bigC{{\hbox{\cmss C}}}
\def\scriptlap{{\kern1pt\vbox{\hrule height 0.8pt\hbox{\vrule width 0.8pt
  \hskip2pt\vbox{\vskip 4pt}\hskip 2pt\vrule width 0.4pt}\hrule height 0.4pt}
  \kern1pt}}
\def\ba{{\bar{a}}}
\def\bb{{\bar{b}}}
\def\bc{{\bar{c}}}
\def\bphi{{\Phi}}
\def\Bigggl{\mathopen\Biggg}
\def\Bigggr{\mathclose\Biggg}
\def\Biggg#1{{\hbox{$\left#1\vbox to 25pt{}\right.\n@space$}}}
\def\n@space{\nulldelimiterspace=0pt \m@th}
\def\m@th{\mathsurround = 0pt}

\begin{titlepage}
\begin{flushright}
{\small OU-HET-1051} 
 \\
\end{flushright}

\begin{center}

\vspace{5mm}


{\Large \bf Notes on islands in asymptotically flat} \\[3pt] 
\vspace{1mm}
{\Large \bf   2d dilaton black holes}  


\vspace{6mm}

\renewcommand\thefootnote{\mbox{$\fnsymbol{footnote}$}}
Takanori Anegawa\footnote{takanegawa@gmail.com} 
and 
Norihiro Iizuka\footnote{iizuka@phys.sci.osaka-u.ac.jp}    

\vspace{3mm}

{\small \sl Department of Physics, Osaka University} \\ 
{\small \sl Toyonaka, Osaka 560-0043, JAPAN}

\end{center}

\vspace{3mm}

\noindent

We study the {\it islands} and the Page curve in the 1+1-dimensional eternal dilaton black hole models. Without islands, the  entanglement entropy of the radiation grows linearly at late time. However with an island, its growth stops at the value of almost twice of the black hole entropy. Therefore an island emerges at the late time, and the entanglement entropy of the radiation shows the Page curve.

\end{titlepage}

\setcounter{footnote}{0}
\renewcommand\thefootnote{\mbox{\arabic{footnote}}}

\newpage

\setcounter{tocdepth}{2}  


\section{Introduction\label{sect:intro}}

Black holes are perhaps most mysterious thermal objects in the universe \cite{Bekenstein:1973ur}. 
In 1974, Hawking discovered that black holes show thermal radiation \cite{Hawking:1974sw}, and argued that the black hole evaporation does not follow unitary evolution; pure states can evolve to mixed states \cite{Hawking:1976ra}. This is because pair particle creations happen near the horizon, and one particle fall into the black hole and the other escapes to asymptotically flat region. These pair particles are highly entangled, and since particles which fall into the black hole cannot escape to asymptotically flat region in semi-classical description, the entanglement entropy between inside and outside black hole will increase eternally. 
However, if unitarity is respected, the entanglement entropy must decrease in the middle of the black hole evaporation and its curve should follow the Page curve \cite{Page:1993wv}. Since Hawking's calculation was 
done in semi-classical approximation, non-perturbative quantum gravity effects, especially breakdown of the locality in quantum gravity as holographic description, are expected to play an important role.

Recently, the idea called \it island \rm was proposed  
to explain the unitary Page curve in the bulk 
\cite{Penington:2019npb,Almheiri:2019psf,Almheiri:2019hni}. 
The proposed formula in \cite{Almheiri:2019hni}  for the ``quantum'' entanglement entropy of the thermal radiation is 
\begin{align}
\label{formula}
S_R={\rm min}\left\{ {\rm ext} \left[\frac{A[\partial I]}{4G_N}+S_{\rm bulk}[{\rm Rad}\cup I]\right]\right\} \,,
\end{align}
where 
$S_{\rm bulk}[{\rm Rad}\cup I]$ implies the entanglement entropy of the matter fields in the bulk spacetime for the region ``Rad'' and $I$.
``Rad'' is the region to which the out-going particles, produced near the horizon, escape under semi-classical limit, and $I$ is so-called the {\it island} region. $\partial I$ is the boundary of $I$ and the first term ${A[\partial I]}/{4G_N}$ is essentially the the Bekenstein-Hawking area-entropy term \cite{Bekenstein:1973ur}. Above formula implies that one should search for all the extremal surfaces for the island boundary $\partial I$, and pick up the one which takes the minimal value for $S_R$.

This formula is based on the prescription of finding the quantum extremal surface for the quantum entanglement entropy \cite{Engelhardt:2014gca}. 
The first term in eq.~\eqref{formula} is Ryu-Takayanagi (RT) extremal surface formula \cite{Ryu:2006bv, Hubeny:2007xt}, the quantitative feature of the entanglement entropy in the bulk in the leading order in $1/G_N$ expansion. 
The second term is the bulk entanglement entropy, which is the quantum corrections of RT formula \cite{Barrella:2013wja, Faulkner:2013ana}. The crucial point of the prescription in \cite{Engelhardt:2014gca} is that one should consider the extremal surface by taking into account this quantum corrections of the bulk.  
In fact, the quantum corrections become important in the process of black hole evaporation. 
At the early time, $I$ is the empty region, therefore the first area term is zero and the second term dominates.  This is the semiclassical Hawking radiation, where the entanglement entropy increases during the black hole evaporation as described above. However once the second term, $S_{\rm bulk}$ grew up to $O(G_N^{-1})$, it becomes the same order as the first area term, then phase transition occurs which causes the change of $I$ \cite{Penington:2019npb,Almheiri:2019psf}.  As the result, suddenly $I$ becomes non-empty region, and the area term increases but the bulk term decreases, in the end the total entropy decreases following the Page curve \cite{Penington:2019npb,Almheiri:2019psf,Almheiri:2019hni}. In this process, the island appears inside the horizon, however, the existence of islands which appear outside the horizon is also verified \cite{Almheiri:2019yqk} by considering eternal black holes. 
Eventually, the causal wedge of the new radiation contains the island $I$. This mechanism is expected to reveal the mystery of the Page curve. In fact, these works show that behaviour of the entanglement entropy is different from the previous one once the phase transition occurs, and do not limitlessly increase.

The emergence of islands are also shown by the replica trick in the Jackiw-Teitelboim (JT) gravity in \cite{Almheiri:2019qdq,Penington:2019kki}. 
As explained there, good analogy to two-dimensional electrodynamics is very useful to understand the emergence of island intuitively: 
In eternal black holes, the distance between left and right asymptotic regions grow linearly in time. In the replica trick, the left and right boundaries of the ``Rad'' region play the role of two opposite charged particles sitting at the left and right asymptotic regions in the two-dimensional electrodynamics.  
Since the distance between the two charged particles is increasing as the time passed, the energy of the electric fields between them also increases, and then a opposite-charged pair particle creation occurs to decreases the energy of electric fields, just as Schwinger mechanism or four-dimensional color confinement. In the replica trick, the positions of the pair created particles above is analogous to the positions of the boundaries of the created island, and the energy of the electric fields is analogous to the entropy.  
Thus, replica method in gravity firmly and securely explains the emergence of the islands at the late time, and why entropy does not grow limitlessly, consistent with the Page curve.

Many works about islands have been done in JT gravity \cite{Jackiw:1984je,Teitelboim:1983ux} coupled to the asymptotically flat region. 
JT gravity is an example of two-dimensional (2d) dilaton gravity, which admits AdS$_2$ holography \cite{Almheiri:2014cka}. 
One can obtain other interesting 2d dilaton gravity models by taking the different form for the potential of the dilaton, see \cite{Grumiller:2002nm} for a nice review. 
Historically, exciting developments of 2d dilaton gravity started from the discovery of the 2d dilaton black hole solutions, first found in \cite{Mandal:1991tz, Witten:1991yr}. 
These 
solutions are different from the solutions of JT gravity in the sense that they are asymptotically flat.
In this paper, we will call these black holes as 2d dilaton black holes.  
The action for these 2d dilaton black holes is motivated by string theory, and it is different from one of the JT gravity.  
The backreaction of the Hawking radiation is taken into account for these 2d dilaton black holes through the matter anomaly in Callan-Giddings-Harvey-Strominger model (CGHS model) \cite{Callan:1992rs}. 
In this paper, we will use the convention of  \cite{Callan:1992rs}. 
%


In order to understand the emergence of the island in more generic setting, in this short paper,  
we study the island in the 2d dilaton black holes.  
We consider the eternal 2d dilaton black holes and we will see that we can construct a Page curve just like \cite{Almheiri:2019yqk}. 
In next section, we calculate the island's position for an eternal black hole metric in 2d dilaton gravity model and reproduce the Page curve.  Section 3 is a short summary, and in Appendix A, we summarize the 2d dilaton black holes. 

{\bf Note added:} After we have finished the calculation in this paper, 
a  new paper \cite{Gautason:2020tmk} appeared on arXiv, which overlaps to this work. That's why we decided to put this result on the arXiv. 

\section{Islands in the 2d dilaton black holes\label{sect:body}}

\subsection{Classical eternal 2d dilaton black holes} 
We consider classical 2d dilaton  black hole background  \cite{Mandal:1991tz, Witten:1991yr, Callan:1992rs}, with CFT matters on this background. The action we have in our mind is  
\begin{align}
S=S_{\rm CGHS}+S_{\rm CFT} \,, 
\end{align}
where $S_{\rm CGHS}$ is given in the appendix, eq.~\eqref{CGHS_action}. 
We mainly consider free CFT in this paper whose central charge is $c$, but in most cases, the detail of the CFT does not matter. 

See the appendix A for the review of the classical 2d dilaton black hole solutions.  
The metric and the dilaton of the eternal, classical 2d dilaton black holes are 
\begin{align}
ds^2 &= - e^{2 \rho} dx^+ dx^- \,, \\
 e^{-2\phi} &=e^{-2\rho}=\frac{M}{\lambda}-\lambda^2x^+x^- \,,
\end{align}
The horizon lies at $x^+x^-=0$. Since this coordinate $(x^+,x^-)$ defines the metric of inside and outside the horizon, we can take pure vacuum state on $\frac{1}{2}(x^++x^-)=0$ slice. These black holes have a temperature $T=\lambda/2\pi$, for an observer at an asymptotically flat region. Following eq.~\eqref{formula}, we will compute the generalized entropy $\frac{\rm Area}{4G_N}|_{\partial I}+S_{\rm bulk}$. 
In this two-dimensional case, the black hole entropy ``area-term'' $\frac{\rm Area}{4G_N}|_{\partial I}$ is given by the dilaton as $2e^{-2\phi}|_{\partial I}=2\left(\frac{M}{\lambda}-\lambda^2x^+x^-\right)|_{\partial I}$ since the black hole entropy is given by $2 e^{-2 \phi}|_{r =r_H}$, where $r=r_H$ is the horizon. 
Since we are interested in the classical black holes, its entropy is expected to be very large. Therefore in this paper, we will mainly consider the large $M/\lambda$ black holes; 
\be
\label{largeMlimit}
\frac{M}{\lambda} \gg 1 \,.
\ee

The asymptotic observer's time is obtained 
by the following change the coordinate as 
\begin{align}
\label{flataymco}
\lambda x^+_R=e^{\lambda \sigma^+_R}\hspace{5.6mm} \,, \quad &\lambda x^-_R=-e^{-\lambda \sigma^-_R}\\
\lambda x^+_L=-e^{-\lambda \sigma^+_L} \,, \quad &\lambda x^-_L=e^{\lambda \sigma^-_L}
\end{align}
$\sigma_{R(L)}$ can describe the right(left) Rindler wedge. And we take $\sigma^{\pm}=t\pm \sigma$. Notice that this $t$, which is the time of the asymptotic observer, has the periodicity along the imaginary axis; $2\pi/\lambda = \beta$, and this agrees with its temperature $T=\lambda/2\pi$. 

In this coordinate, the metric becomes
\begin{align}
\label{asymflat_t_coord}
&-\frac{dx^+dx^-}{\frac{M}{\lambda}-\lambda^2x^+x^-}=-\frac{d\sigma_R^+d\sigma_R^-}{\frac{M}{\lambda}e^{\lambda(\sigma_R^--\sigma_R^+)}+1}, \\
&-\frac{dx^+dx^-}{\frac{M}{\lambda}-\lambda^2x^+x^-}=-\frac{d\sigma_L^+d\sigma_L^-}{\frac{M}{\lambda}e^{\lambda(\sigma_L^+-\sigma_L^-)}+1}
\end{align}
By taking the limit $\sigma_R^+ - \sigma_R^- \to \infty$ or $\sigma_L^- - \sigma_L^+ \to \infty$, this metric will approach the flat one.

\subsection{Entanglement entropy without islands} 

\begin{figure}
\begin{center}
\includegraphics[bb={0 0 300 160}]{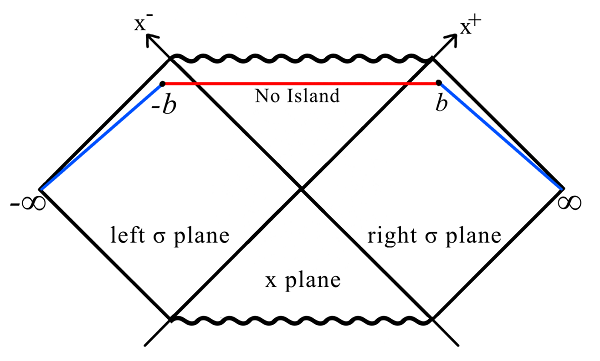}
\caption{We consider the entropy of the interval $[-\infty,-b]_L\cup[b,\infty]_R$. This entropy, which is the entanglement entropy between pair created particles near the horizon, increases linearly and eternally in time at the late time.  
If unitarity is hold, the fine-grained entropy cannot increase eternally and it must to reach to a fixed value at the late time, which is contradicted with the linear growth.} 
\label{noislandpic}
\end{center}
\end{figure}

Without island, the first area term in eq.~\eqref{formula} vanishes. Therefore  
we just need to calculate bulk entropy. We will take the intervals $[-\infty,-b]_L\cup [b,\infty]_R$ in $\sigma$ coordinates to lie on the non-zero time slice $t$. as Fig.~\ref{noislandpic}.  The values of these coordinates are given by 
\begin{align}
\sigma_R^{\pm}=t\pm b \,, \quad \sigma_L^{\pm}=t\mp b \,.
\end{align}
The bulk state is taken to be vacuum in $(x^+,x^-)$ coordinate. Then, the entanglement entropy is given by the universal formula as
\begin{align}
S=\frac{c}{6}\log\left( \frac{|x_{RL}^+x_{RL}^-|}{\sqrt{\frac{M}{\lambda}-\lambda^2x_R^+x_R^-}\sqrt{\frac{M}{\lambda}-\lambda^2x_L^+x_L^-}}\right)
\end{align}
where $x^\pm_{RL} = x^\pm_R - x^\pm_L$. 
The denominators comes from the warp factor of the metric. 
%
In terms of  $\sigma_R^{\pm}=t\pm b,\ \ \sigma_L^{\pm}=t\mp b$, the entropy is
\begin{align}
S_{\rm no\ island}=\frac{c}{6}\log\left( \frac{2^2\cosh^2{\lambda t}}{\lambda^2\left( \frac{M}{\lambda}e^{-2\lambda b}+1\right)}\right)\simeq \frac{c\lambda}{3}t+\cdots\ \ {\rm for}\ t\gg \lambda^{-1}
\end{align}

This grow linearly. This linear growth have roots in entanglement between particles which are produced by pair creations near the horizon. This linear growth appears only when we use time of the asymptotic observer. If we use any other coordinate, the entropy does not usually increase linearly.  
This eternal growth contradicts with the finiteness of the fine-grained entropy, which is $O(M/\lambda)$. 
Therefore especially in the late time where 
\be
\label{def_of_latetime}
\frac{M}{ \lambda} \ll c \lambda t \to \infty
\ee
we have contradiction with the Page curve and this corresponds to the information paradox. From now on, what we mean the late time is  eq.~\eqref{def_of_latetime}.

\subsection{Entanglement entropy with islands}

\begin{figure}
\begin{center}
\includegraphics[bb={0 0 300 160}]{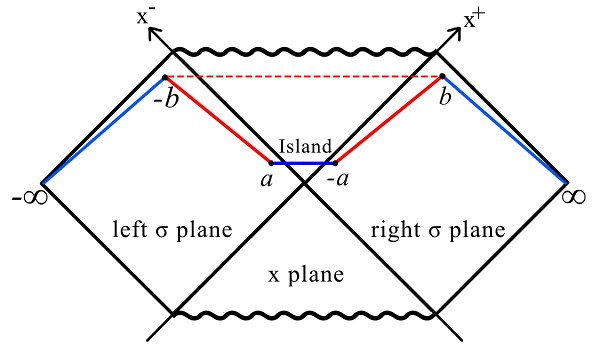}
\caption{We add the interval $[a_L,-a_R]$ in $\sigma$ coordinates. This island region lies on the non-zero time $t'$ slice. After the Page time $t_{\rm Page}$, this entropy becomes less than the entropy of no-island configuration. Thus, a phase transition occurs.}
\label{islandpic}
\end{center}
\end{figure}

Now we will calculate the entanglement entropy with islands. 
We will calculate an entropy when we take the intervals $[-b,a]_L\cup [-a,b]_R$ in $\sigma$ coordinates to lie on the non-zero time slice $t'$ as Fig.~\ref{islandpic}. In this case, islands region is $[a,\infty]_L\cup [-\infty,-a]_R$. Then, the positions of the boundaries of the island is given by 
\be
\sigma_R^{\pm}=t'\mp a \,,  \quad \sigma_L^{\pm}=t'\pm a 
\ee

Since at the late time $t \to \infty$, satisfying eq.~\eqref{def_of_latetime}, the entropy calculated in previous subsection shows the significant deviations from the Page curve, we focus on the late time behaviour; $t \to \infty$.  
In this case, 
the leading term of this von Neumann entropy of the CFT matter can be well approximated and given by twice of the one in the single interval on the right side \cite{Almheiri:2019yqk}, 
\begin{align}
S_{\rm island}=4\left( \frac{M}{\lambda}+e^{-2\lambda a}\right)+\frac{c}{3}\log\left( \frac{\left(e^{\lambda(t+b)}-e^{\lambda(t'-a)}\right)\left(e^{-\lambda(t-b)}-e^{-\lambda(t'+a)}\right)}{\lambda^2 \sqrt{\frac{M}{\lambda}+e^{2\lambda b}}\sqrt{\frac{M}{\lambda}+e^{-2\lambda a}}}\right)
\end{align}
See the appendix B for an explicit justification of this approximation for the free fermion case, where justification of this approximation can be seen as far as  $t, t' \to + \infty$, {\it i.e.,} $t$ and $t'$ are large compared with $a$ and $b$.

Extremizing $S_{\rm island}$ with respect to $t'$, we obtain 
\begin{align}
\partial_{t'}S_{\rm island}
&=\frac{c\lambda}{3}\frac{e^{\lambda(b-a)}\left(-e^{-\lambda(t-t')}+e^{\lambda(t-t')}\right)}{\left(e^{\lambda(t+b)}-e^{\lambda(t'-a)}\right)\left(e^{-\lambda(t-b)}-e^{-\lambda(t'+a)}\right)}=0  \\
& \qquad \qquad \qquad \Rightarrow  \quad t = t'  \,.
\end{align}
Then under $t' = t$, by extremizing $S_{\rm island}$ with respect to $a$, we have 
\begin{align}
\hspace{-1mm}&\partial_{a}S_{\rm island} 
= -8\lambda e^{-2\lambda a}+\frac{2c\lambda }{3}\frac{1}{e^{\lambda (b+a)}-1}-\frac{c}{6}\frac{\left( -2\lambda \right)}{\frac{M}{\lambda}e^{2\lambda a}+1}=0
\end{align}
To simplify this equation, we solve this equation under the ansatz that 
\be
\label{quadeqapproximation}
\frac{M}{\lambda}e^{2\lambda a}+1\simeq \frac{M}{\lambda}e^{2\lambda a}  
\ee
and then later, we check its consistency. 
%
Setting $X=e^{\lambda a}$, 
this equation reduces to 
\begin{align}
\frac{c}{6}X^2-\left(2-\frac{c\lambda}{12M} \right)e^{\lambda b}X+\left(2-\frac{c\lambda}{12M} \right)=0 \,,
\end{align}
and its solution is
\begin{align}
X = e^{\lambda a} 
&=\frac{3}{c}\left(2-\frac{c\lambda}{12M} \right)e^{\lambda b} \left(1\pm\sqrt{1-\frac{2c}{3}e^{-2\lambda b}\frac{1}{\left(2-\frac{c\lambda}{12M} \right)}}\right) \,.
\label{Xsoln}
\end{align}
Since the island boundary must extremize the generalized entropy, the value of $a$ must be real and 
inside of the square root of above must be positive.  
Mathematically we can consider two possibilities for that; 
(A) $M/\lambda \gg c$ and $e^{2 \lambda b} \gg c$, or  
(B) $M/\lambda \ll c $.  
Since we are interested in the classical black holes where its entropy is large as eq.~\eqref{largeMlimit}, 
case (B) is not satisfied unless we consider extremely large $c$ limit. In this paper, 
we will not consider the case (B), instead we focus on the parameter range satisfying (A) above.   
This condition implies that $c$ is not parametrically large, and from $e^{2 \lambda b} \gg c$, we take large $b$. 
In summary the parameter range we consider is  
\be
\label{ouparameterrange1}
\frac{M}{\lambda} \gg \lambda b \gg c  \,.
\ee
where $c$ takes the fixed value. 
The reason why we consider the parameter range $\frac{M}{\lambda} \gg \lambda b $ becomes clear later. Furthermore, we require that $\sigma = b$ is near the asymptotically flat region, then we need additionally the condition 
\be
\label{ouparameterrange2}
e^{2 \lambda b} \gg \frac{M}{\lambda} \,, 
\ee
as seen from the metric eq.~\eqref{asymflat_t_coord}, even though we require $\frac{M}{\lambda} \gg \lambda b $.  
With these, we consider the late time as eq.~\eqref{def_of_latetime}.

In this limit, eq.~\eqref{Xsoln} becomes 
\begin{align}
X& = e^{\lambda a} =\frac{3}{c}\left(2-\frac{c\lambda}{12M} \right)e^{\lambda b} \left(1\pm\left( 1-\frac{c}{3}e^{-2\lambda b}\frac{1}{\left(2-\frac{c\lambda}{12M} \right)}-O(e^{-4\lambda b})\right)\right)\nonumber \\
&\simeq 
\begin{cases}
    \frac{6}{c}\left(2-\frac{c\lambda}{12M} \right)e^{\lambda b} &({\rm plus\ sign})\\
    e^{-\lambda b}+O(e^{-3\lambda b}) &({\rm minus\ sign})
  \end{cases}
\end{align}
For later convenience, we keep the subleading term when we take minus sign. 
We can conclude the position of island is 
\begin{align}
\label{islandsol}
a\simeq 
\begin{cases}
    b &({\rm plus\ sign})\\
    -b+O(e^{-2\lambda b}) &({\rm minus\ sign})
  \end{cases}
\end{align}

(i) If we take $a\simeq b$, then the entropy becomes
\begin{align}
S_{\rm island(i)}&\simeq 4\left( \frac{M}{\lambda}+e^{-2\lambda b}\right)+\frac{c}{3}\log\left( \frac{\left(e^{\lambda b}-e^{ - \lambda b}\right)^2}{\lambda^2 \sqrt{\frac{M}{\lambda}+e^{2\lambda b}}\sqrt{\frac{M}{\lambda}+e^{-2\lambda b}}}\right)\nonumber \\
&\simeq 4\left( \frac{M}{\lambda}+e^{-2\lambda b}\right) + \frac{2c}{3}  \lambda b-\frac{c}{6}\log \left( \frac{M}{\lambda}+e^{2\lambda b}\right)-\frac{c}{6}\log \left( \frac{M}{\lambda}+e^{-2\lambda b}\right)\nonumber \\
&= 2S_{BH} + O\left(\frac{c \lambda b}{\frac{M}{\lambda}}\right)
\end{align}
under the parameter range, eq.~\eqref{ouparameterrange1} and eq.~\eqref{ouparameterrange2}. Here $S_{BH} = 2 e^{ - 2 \phi}|_{r =r_H}= 2M/\lambda$. 
The ansatz eq.~\eqref{quadeqapproximation} is certainly satisfied under these parameter range.  

(ii) The case $a\simeq -b$, cannot give the minimal value of the entropy in our parameter range. This can be seen by the followings; 
the entropy for the case of $a\simeq -b$  becomes
\begin{align}
\label{isvab}
S_{\rm island(ii)}&=4\left( \frac{M}{\lambda}+e^{2\lambda b}\right)+\frac{c}{3}\log\left( \frac{\left(e^{\lambda b}-e^{-\lambda a}\right)^2}{\lambda^2 \sqrt{\frac{M}{\lambda}+e^{2\lambda b}}\sqrt{\frac{M}{\lambda}+e^{-2\lambda a}}}\right)\nonumber \\
&\simeq 2S_{BH}+4e^{2\lambda b}\gg S_{\rm island(i)} \,.
\end{align} 
under the parameter range, eq.~\eqref{ouparameterrange1} and eq.~\eqref{ouparameterrange2}. 
Here the $\log$ term seems to diverge, if we naively plug in $a = -b$. However $a$ is not strictly $-b$, and $e^{\lambda b}-e^{-\lambda a}\sim O(e^{-\lambda b})\ll 1$. Thus this log divergence gives at most $\simeq -2\lambda b$. 

Needless to say, in the case $a\simeq -b$ the ansatz eq.~\eqref{quadeqapproximation} is not satisfied under the parameter range eq.~\eqref{ouparameterrange1} and eq.~\eqref{ouparameterrange2}. 
However, the point that $a\simeq -b$ cannot give the minimal value of the entropy in our parameter range, is independent of whether $a\simeq -b$ extremize the generalized entropy or not. 
Therefore $a\simeq -b$ is excluded as an island boundary. 

\begin{figure}
\begin{center}
\includegraphics[width=84mm,bb={10 0 300 260}]{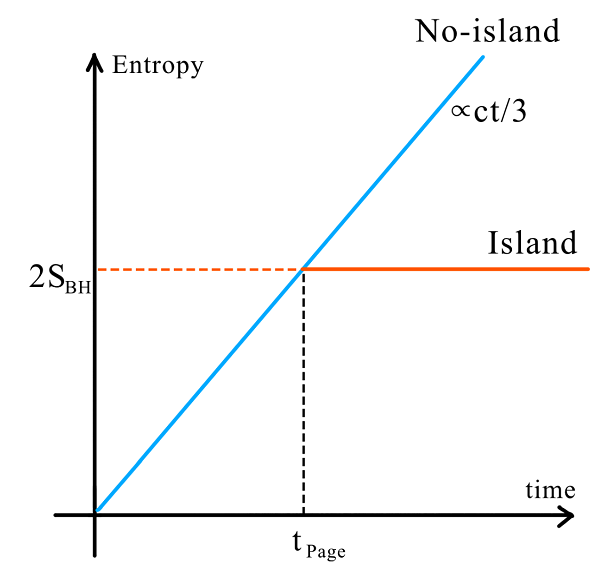}
\caption{The Page curve. The blue line is the entropy without islands, which increases eternally. After the Page time $t_{\rm Page} \simeq 6 \beta S_{BH}/2 \pi c$, the one without islands becomes greater than the one with an island. Thus phase transition occurs, and the entropy reaches an constant value, which is approximately twice of the black hole entropy.} 
\label{pagecurve}
\end{center}
\end{figure}

In summary,  no-island phase dominates in the early time as Fig.~\ref{pagecurve}. 
However, $S_{\rm no\ island}$ increase linearly and limitlessly without islands. This is in conflict with the maximal fine-grained entropy $S_{\rm max}=2S_{BH}$, which is entropy of two black holes. 
In fact, after the Page time $t = t_{\rm Page}$ where 
\be
t_{\rm Page} \eqsim \frac{6 S_{BH}}{c \lambda} = \frac{6 \beta S_{BH}}{2 \pi c}= \frac{12 M}{c \lambda^2}  
 \,, 
\ee 
the island phase comes to dominate, the entropy is roughly $2S_{BH}$ in our parameter range with small corrections.  
The island boundary is $\sigma \simeq - b$, which is outside of the horizon $\sigma_H = - \infty$.

\section{Short summary}
\label{sec:Summary}
In this rather short paper, we studied the emergence of an island in 1+1 dimensional eternal 2d dilaton black holes. 
Although the entropy increases eternally after the Page time without islands, we found an island appears  
outside the horizon after the Page time. 
The island makes the radiation entropy finite,  which is twice of the black hole entropy, and it is  an upper bound to this radiation entropy as shown in Fig.~\ref{pagecurve}.

Our calculation was done under some appropriate approximations, in particular, we consider the late time eq.~\eqref{def_of_latetime} with 
the parameter range eq.~\eqref{ouparameterrange1} and eq.~\eqref{ouparameterrange2}. 
Studying the behaviour of the islands in more generic parameter range is definitely worth investigating.

Finally to study the various natures of the islands, it is interesting the behaviour of islands in variety of other models, see \cite{Grumiller:2002nm}. 
It is also interesting to study them in higher dimensions.  
For examples, see \cite{Almheiri:2019psy, Hashimoto:2020xxx}. 

\section*{Acknowledgement} 
We would like to thank Koji Hashimoto,  Yoshinori Matsuo, and Kotaro Tamaoka for discussions. 
The work of N.I. was supported in part by JSPS KAKENHI Grant Number 18K03619.


\appendix

\section{Review of 2d dilaton black holes}
\label{AppA}
In this appendix, we review the 2d dilaton black hole solutions \cite{Mandal:1991tz, Witten:1991yr, Callan:1992rs}. 

Both JT gravity \cite{Jackiw:1984je, Teitelboim:1983ux} and the CGHS model \cite{Callan:1992rs}
are two-dimensional dilaton gravity models. Their bulk action can be summarized in the following model \cite{Grumiller:2002nm}; 
\be
\label{A1}
S = \frac{1}{16 \pi G_N} \int d^2x \sqrt{-g} \left[ \tilde{\Phi}  \left( R + K(\tilde{\Phi}) ( \nabla \tilde{\Phi} )^2 - 2 V(\tilde{\Phi})  \right) \right] 
\ee
where $K(\tilde{\Phi})$ and $V(\tilde{\Phi})$ are 
\begin{align}
&K = 0 \,, \quad V = - \lambda^2 \quad \qquad \,\,\, \mbox{(for JT gravity)} \\
&K = \frac{1}{\tilde{\Phi}^2} \,, \quad V = - 2 \lambda^2   \quad \,\,\,\,\mbox{(for CGHS model)}
\end{align}
Here $\lambda$ sets the length scale of the cosmological constant. 
JT gravity admits AdS$_2$. 
For the CGHS model, by the field redefinition as $\tilde{\Phi} \equiv e^{-2 \phi}$, above action reduces to the following form \cite{Callan:1992rs}:
\be
\label{CGHS_action}
S_{\rm CGHS} = \frac{1}{16 \pi G_N} \int d^2x \sqrt{-g} \left[ e^{- 2 \phi}  \left( R + 4 \left( \nabla \phi  \right)^2 + 4 \lambda^2  \right) \right] 
\ee 
By solving the equations of motions from this action, one can obtain the vacuum black hole solutions, first found in \cite{Mandal:1991tz, Witten:1991yr}\footnote{Furthermore, by taking into account conformal anomaly one can calculate the Hawking radiation and its backreaction \cite{Callan:1992rs}, which we will not consider in this paper.  
There are many relevant papers and good reviews about this model,  
see for examples, \cite{Giddings:1992fp,Fiola:1994ir,Strominger:1994tn,Almheiri:2013wka}.}.

In terms of two-dimensional light-cone coordinates $x^{\pm}=x^0\pm x^1$, and in the conformal gauge,  
the vacuum black hole solutions \cite{Mandal:1991tz, Witten:1991yr, Callan:1992rs}  are,  
\begin{align}
\label{congauge}
ds^2 = &- e^{2 \rho} dx^+ dx^- 
\end{align}
as 
\begin{align}
\label{solofphi}
e^{-2\phi}=e^{-2\rho}=\frac{M}{\lambda}-\lambda^2x^+x^- \,,
\end{align}
where $M > 0$ is a parameter of the solution but it essentially represents the mass of the black hole. 
By setting $M=0$, we have the flat space time according to ordinary expectations. Then the value of the dilaton is
\begin{align}
\phi=-\frac{1}{2}\log(-\lambda^2 x^+x^-)=-\frac{1}{2}\lambda(\sigma^+-\sigma^-)=-\lambda \sigma
\end{align} 
by using eq.~\eqref{flataymco}. This dilaton is proportional to $\sigma$. This is the linear dilaton vacuum.

To show where the horizon lies, we set
\begin{align}
t=\frac{1}{2\lambda}\log\left(-\frac{x^+}{x^-}\right),\ \ r=\frac{1}{2\lambda}\log\left(\frac{M}{\lambda}-\lambda^2x^+x^- \right)
\end{align}
This gives for the new metric
\begin{align}
\label{likeSchw}
ds^2=-\left( 1-\frac{M}{\lambda}e^{-2\lambda r}\right)dt^2+\frac{dr^2}{1-\frac{M}{\lambda}e^{-2\lambda r}} \,, 
\end{align}
which is exactly like the Schwarzschild metric, admitting asymptotically flat spacetime. Thus, the horizon $r=r_H$ is given by 
\be
\label{thisishorizon}
r_H = \frac{1}{2\lambda}\log\frac{M}{\lambda} \quad \Leftrightarrow \quad x^+x^-=0 \quad \mbox{(horizon)}
\ee
Ricci scalar in conformal gauge eq.~\eqref{congauge} is given by
\begin{align}
R = 8   e^{- 2 \rho}  \partial_+ \partial_- \rho   =\frac{4M\lambda}{\frac{M}{\lambda}-\lambda^2x^+x^-}
\end{align}
which is divergent at $x^+x^-=M/\lambda^3$ (singularity).

We can show this black hole spectrum obeys Planck distribution \cite{Giddings:1992ff}. From the averaged energy of black body radiation in two dimensions, we can compute the Hawking temperature as 
\begin{align}
&\int _0^{\infty}\frac{dp}{2\pi} \frac{|p|}{e^{|p|/T}-1}=\frac{\pi}{12}T^2=\braket{T_{--}}=\frac{\lambda^2}{48\pi}  \quad 
 \Rightarrow \quad   T=\frac{\lambda}{2\pi} \,.
\end{align}
This temperature can be confirmed from the metric eq.~\eqref{likeSchw} with Euclidean time periodicity 
\be
\beta = \frac{1}{T} = \frac{2 \pi}{\lambda} \,,
\ee 
by removing the conical singularity at the horizon as a usual technique. 
The entropy of the black hole is given by the dilaton value at the horizon 
\begin{align}
S_{\rm BH} =  2 e^{-2\phi}|_{r =r_H} = \frac{2M}{\lambda} \,. 
\end{align}


\section{Approximation of the entropy formula}
\label{AppB}
In this appendix, we approximate the entropy formula of two disjoint intervals. 
The explicit formula for free Dirac fermion is given by \cite{Casini:2005rm}
\begin{align}
S_{\rm fermions}=\frac{c}{6}\log\left[ \frac{|x_{21}x_{32}x_{43}x_{41}|^2}{|x_{31}x_{42}|^2\Omega_1 \Omega_2 \Omega_3 \Omega_4}\right]
\end{align}
where two disjoint interval is set $[x_1,x_2]\cup [x_3,x_4]$ and $\Omega$ is a warp factor $ds^2=-\Omega^{-2}dx^+dx^-$.  
We choose $x_1$/$x_3$ representing the right/left boundary of the island, and $x_2$/$x_4$ representing the right/left boundary of the Rad region. 
We compute an explicit condition for approximations which we used in Section 2.3.

In Section 2.3 we approximate this entropy as the sum of entropies of single interval. 
The claim is that in the limit $a, b \ll t ,  t'  \to + \infty$, this approximation can be justified. 
To see this, we rewrite this explicit entropy as 
\begin{align}
S_{\rm fermions}=\frac{c}{6}\log\left[ \frac{|x_{21}|^2}{\Omega_1\Omega_2}\right]+\frac{c}{6}\log\left[ \frac{|x_{43}|^2}{\Omega_3\Omega_4}\right]+ \frac{c}{6}  \log\left[\frac{|x_{32}x_{41}|^2}{|x_{31}x_{42}|^2}\right]
\end{align}
The third term is negligible if both $t$ and $t'$ goes to infinity, since then  
\begin{align}
\frac{|x_{32}x_{41}|^2}{|x_{31}x_{42}|^2}=\frac{(e^{\lambda (t+b+a)}+e^{-\lambda t'})^2(e^{\lambda (t'-b-a)}+e^{-\lambda t})^2}{(e^{-\lambda t'}+e^{\lambda t'})^2(e^{-\lambda t}+e^{\lambda t})^2} \to 1
\end{align}
In fact, the islands appearing at late time of section 2.3 satisfies $a + b \ll t \simeq t'  \to + \infty$, consistent with our approximation.


\end{document}